Article

# Single-photon counting pixel detector for soft X-rays

Check for updates

Filippo Baruffaldi[1,4], Anna Bergamaschi[1] ✉, Maurizio Boscardin[2], Martin Brückner[1], Tim A. Butcher[1,3], Maria Carulla[1], Matteo Centis Vignali[2], Roberto Dinapoli[1], Simone Finizio[1], Erik Fröjdh[1], Dominic Greiffenberg[1], Aldo Mozzanica[1], Giovanni Paternoster[2], Nicholas W. Phillips[1,5], Jörg Raabe[1], Bernd Schmitt[1] & Jiaguo Zhang[1]

Soft X-ray experiments at synchrotron light sources are essential for a wide range of research fields. However, commercially available detectors for this energy range often cannot deliver the necessary combination of quantum efficiency, signal-to-noise ratio, dynamic range, speed, and radiation hardness within a single system. While hybrid detectors have addressed these challenges effectively in the hard X-ray regime, specifically with single photon counting pixel detectors extensively used in high-performance synchrotron applications, similar solutions are desired for energies below 2 keV. In this work, we introduce a single photon counting hybrid pixel detector capable of detecting X-ray energies as low as 550 eV, utilizing the internal amplification of Low Gain Avalanche Diode (LGAD) sensors. This detector is thoroughly characterized in terms of Signal-to-Noise Ratio and Detective Quantum Efficiency. We demonstrate its capabilities through ptychographic imaging at MAX IV 4th-generation synchrotron light source at the Fe $L_3$-edge (707 eV), showcasing the enhanced detection performance of the system. This development sets a benchmark for soft X-ray applications at synchrotrons, paving the way for significant advancements in imaging and analysis at lower photon energies.

Several imaging techniques rely on low energy X-rays, including Scanning Transmission X-ray Microscopy (STXM), full-field Transmission X-ray Microscopy (TXM), anomalous diffraction, X-Ray Magnetic Circular and Linear Dichroism (XMCD and XMLD) and Coherent Diffractive Imaging (CDI)[1–5]. Working at the absorption edges within the soft X-ray range between 250 eV and 2 keV offers enhanced sensitivity to specific chemical species. For instance, the K-edges of light elements such as C, O, N, F, and P, which are common in many organic and biological systems[6,7], as well as the L-edges of 3d transition metals like Cu, Ni, and Fe, as well as the M-edges of rare earth elements, which are crucial for research on magnetic, ferroelectric, and electrode materials[8–12], fall within this range.

In the early 2000s, hybrid photon-counting pixel detectors with silicon sensors revolutionized tender and hard X-ray applications at synchrotrons, replacing previously used Charge-Coupled Devices (CCDs). These detectors greatly advanced a range of experimental techniques, including macromolecular crystallography (MX), small-angle scattering (SAXS), CDI, and powder diffraction[13,14], thanks to their superior dynamic range and the absence of read-out noise. Moreover, their high frame rate enabled methods such as fine $\varphi$-sliced and time-resolved crystallography[15], as well as scanning microscopy techniques such as ptychography[16].

To date, the use of single photon counting detectors has been largely limited to hard and tender X-ray energy ranges. The lowest reported energy at which Single Photon Counting (SPC) hybrid pixel detectors have been used is 1.75 keV[17]. To the best of our knowledge, position-sensitive SPC detectors for lower energy photons have not yet been available. The primary reason is that the small signals produced by single photons are nearly indistinguishable from the electronic noise of the detector[18].

Hybrid detectors consist of a sensor, which absorbs radiation and converts it into electric charge, connected to an Application Specific Integrated Circuit (ASIC) via high-density flip-chip bonding on a pixel-by-pixel basis, as illustrated in Fig. 1a. Sensors must be optimized to maximize the fraction of incident photons that are absorbed, converted into electric charge, and successfully collected. This is referred to as the Quantum Efficiency (QE). However, standard silicon sensors are typically not ideal for achieving high QE for low-energy photons. This is due to their limited ability to collect charge carriers generated just below the entrance window of the sensor. The primary advantage of the hybrid approach is that the sensor and readout electronics can be optimized independently, with few technological constraints, as demonstrated in this work. This is a significant advantage over monolithic detectors, in which the sensor and readout electronics are integrated onto the same ASIC[19].

[1] Center for Photon Science, Paul Scherrer Institut, Villigen PSI, Switzerland. [2] Fondazione Bruno Kessler, Trento, Italy. [3] Max Born Institute for Nonlinear Optics and Short Pulse Spectroscopy, Berlin, Germany. [4] Present address: Dectris AG, Baden-Dättwill, Switzerland. [5] Present address: CSIRO, Mineral Resources, Clayton, VIC 3168, Australia. ✉ e-mail: anna.bergamaschi@psi.ch

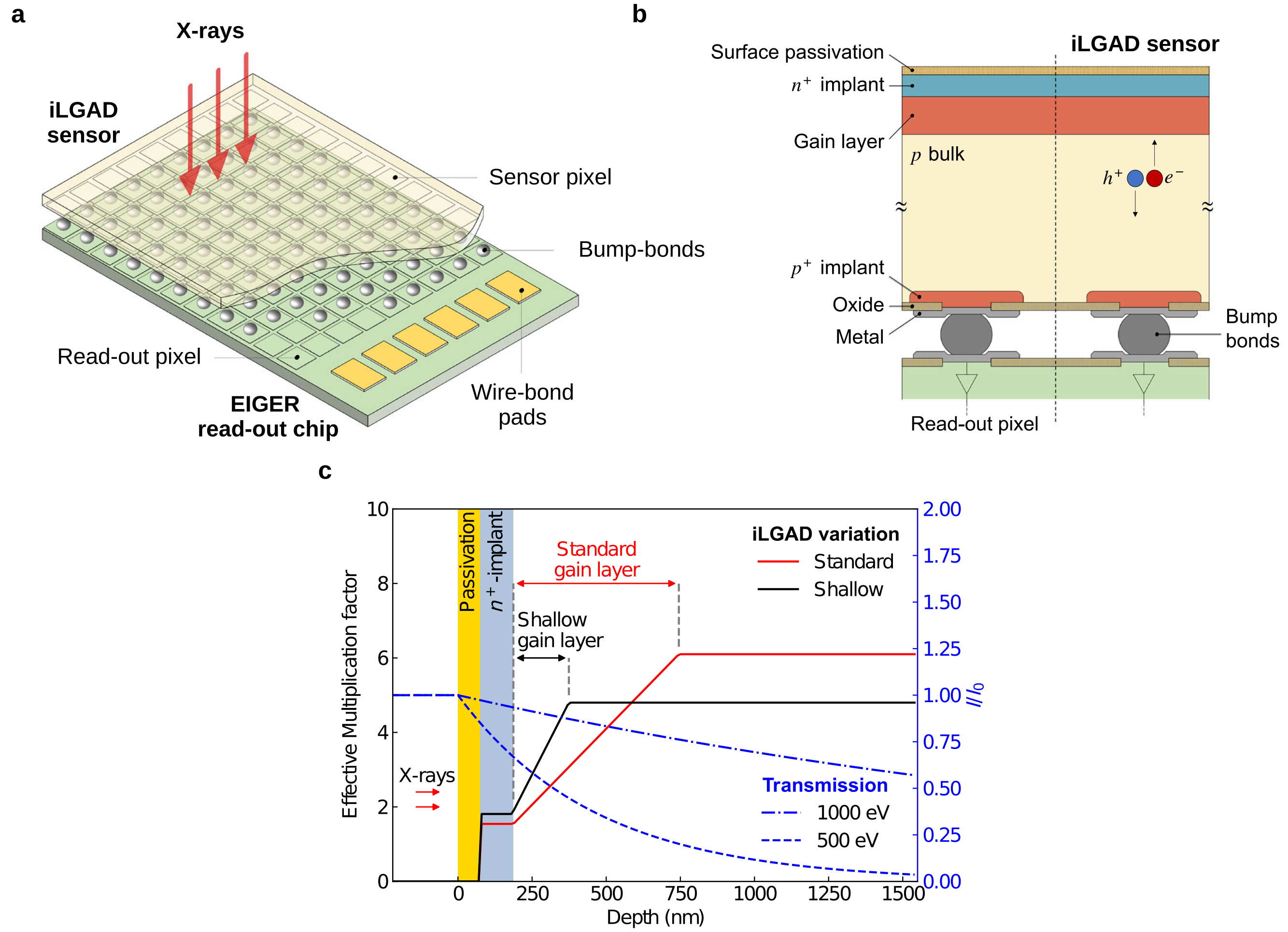

**Fig. 1 | Hybrid pixel detector and LGAD sensors. a** Layout of the hybrid pixel detector used. The iLGAD sensor, where X-rays are converted into electric charge, is connected pixel-by-pixel to the EIGER readout electronics using bump bonding. The readout chip is biased, controlled, and read out through wire-bond pads. **b** Sketch of the cross-section of the inverse-LGAD sensor. X-rays enter from the top through a thin entrance window composed of passivation and a shallow $n^+$ doping layer. The $p^+$ gain layer forms a junction with a high electric field, enabling charge multiplication via impact ionization. The signal is induced in the $p^+$ pixels, which are connected to the readout electronics using bump bonding. **c** Multiplication factor $M$ as a function of photon absorption depth for standard (red) and shallow (black) iLGAD variations[45]. The blue lines and secondary $y$-axis axis represent the transmission of 500 eV (dashed) and 1000 eV (dash-dotted) photons.

Monolithic detectors with an optical entrance window generally have higher QE in the soft X-ray range compared to hybrid detectors and exhibit lower electronic noise thanks to their low input capacitance[20]. Consequently, low-energy applications often use CCDs[21–23] and CMOS sensors[24,25], despite their limitations in frame rate, dynamic range, and radiation hardness. Recent developments, however, aim to address and overcome these drawbacks[26–28]. Ptychography in the soft X-ray energy range is mostly performed using fast CCD devices[29,30] or CMOS sensors[28,31–34].

Charge integrating hybrid detectors with low electronic noise can operate at lower energies, down to a few hundred eV[35] or even into the EUV range without single photon resolution[36,37], but they come with the drawback of limited dynamic range and more complex operation. The development of hybrid single photon counting detector systems for soft X-rays, with performance comparable to those currently achieved at higher energies, could significantly improve many experimental techniques currently limited by available detector performance[1].

This study presents the results achieved by combining the EIGER single photon counting readout chip[38] with Low Gain Avalanche Diode (LGAD) sensors optimized for soft X-ray detection[39]. The system takes advantage of the hybrid architecture's flexibility, substituting the standard silicon p-in-n pixel sensor with inverse LGAD devices and using their internal signal amplification to detect single low-energy photons[18,40]. To the best of our knowledge, this work marks the first demonstration of a single photon counting pixel detector capable of detecting X-rays down to 550 eV. We present the performance of the system along with initial experimental results in ptychographic imaging at the Fe $L_3$-edge at 707 eV.

## Results

### Detector description

LGAD sensors are based on an $n$-in-$p$ silicon junction and incorporate an additional layer, moderately doped with the same polarity as the substrate. This region, typically created via ion implantation and referred to as the gain implant, is located just beneath the shallow surface junction. When fully depleted, this $p^+$ region exhibits an electric field high enough to enable the generation of secondary charge carriers by impact-ionization when electrons or holes travel through it. While this concept has been widely employed in avalanche photodiodes (APDs) and single-photon avalanche detectors (SPADs), LGADs are specifically designed to target a low multiplication factor $M$ of ~5–20, avoiding dark counts while allowing segmentation. Originally proposed in the early 2010s to achieve high timing resolution in tracking detectors for high-energy physics experiments[41–43], this effect can also be used to amplify signals from low-energy photons, boosting their signal above the noise of the readout electronics[18], similar to the noise reduction achieved in Electron-Multiplying CCDs (EM-CCD)[44]. However, compared to the EM-CCD, the multiplication happens before charge collection and storage instead of at the time of readout, without affecting the frame rate. The LGAD sensors used in this study have a uniform gain layer directly beneath the entrance window, as shown in Fig. 1b, enabling a 100% fill-factor, while the fine pixel segmentation is fabricated on the opposite side of the sensor, allowing the hybridization with the read-out electronics. These sensors are referred to in the literature as inverse-LGADs (iLGADs)[40].

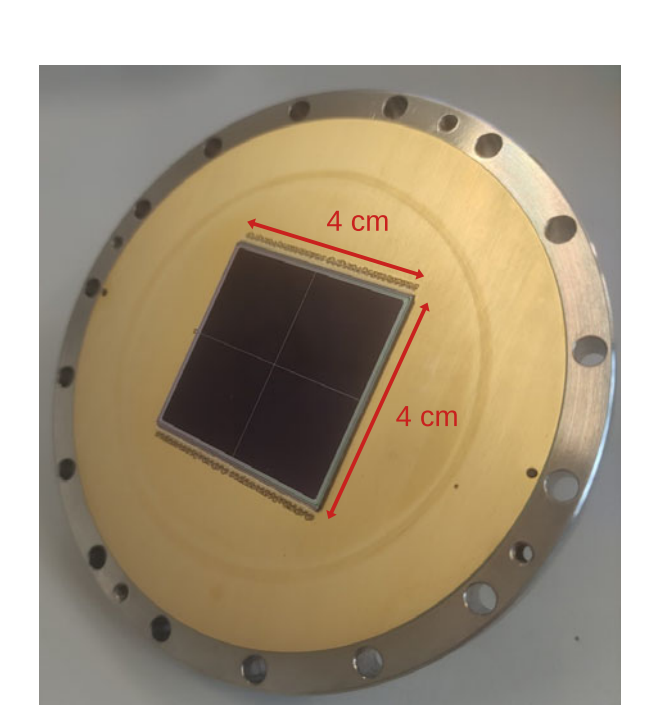

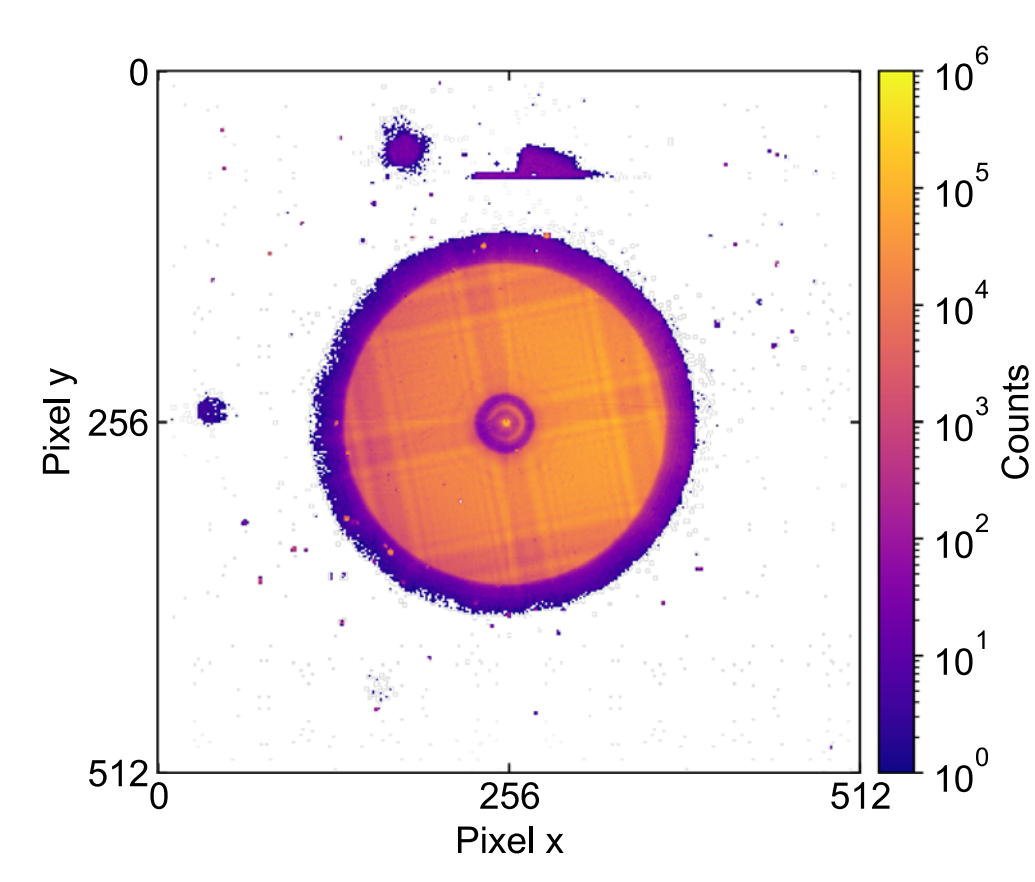

**Fig. 2 | EIGER-iLGAD.** **a** Picture of a 4 × 4 $cm^2$ inverse-LGAD sensor bump-bonded to an EIGER ASIC, mounted on a vacuum-compatible flange. **b** Image of the far-field diffraction from a Frenel Zone Plate focusing optics projected on the detector surface acquired with a 550 eV photon beam (single frame with 1 s exposure time). Additional diffraction and scattered X-rays are visible in the background. The set threshold corresponds to ~480 eV. Fewer than 1% noisy pixels have been masked out.

Maximum charge multiplication occurs when an incident photon is absorbed in the sensor bulk, behind the gain layer. The generated electrons then drift toward the entrance window, crossing the gain layer and producing secondary electrons and holes that drift towards the entrance window and the pixel side, respectively. When photons are absorbed at shallower depths, i.e., in the $n^+$ region or within the gain layer, charge multiplication is initiated either partially or entirely by the holes crossing the gain layer and drifting toward the pixel contacts. Since the ionization coefficient of holes is lower than that of electrons in silicon, this results in a lower effective multiplication factor $M$, as illustrated in Fig. 1c, which shows $M$ as a function of the depth of the photon absorption. To address this, iLGAD variations with a shallower gain layer were investigated. In the standard design, the gain layer extends to a depth of 800 nm, while in the shallow design, it reaches about half of that depth, thus increasing the fraction of electron-initiated events[45].

To overcome the poor QE of conventional silicon sensors in the soft X-ray range, a custom entrance window process has been developed[39]. This optimized entrance window is based on a customized doping profile and includes a thin $SiO_2$ and $Si_3N_4$ passivation layer on the surface, enhancing charge collection efficiency and achieving a QE of up to 55% at 250 eV, limited only by the thickness of the passivation layer[45,46]. In comparison, conventional sensors typically achieve a QE of less than 5% at this energy[17].

These LGAD sensors are combined with the 75 μm pitch single photon counting EIGER ASIC, widely used in hard X-ray applications worldwide[47–51]. Additional details about the EIGER ASIC are available in "Methods" section. The system comprises 2 × 2 ASICs bump-bonded to a 275 μm thick iLGAD sensor with 512 × 512 pixels, covering an area of 4 × 4 $cm^2$, as shown in Fig. 2a. The sensors used in this study, fabricated by Fondazione Bruno Kessler (FBK, Trento, Italy), to the best of our knowledge, are the largest LGAD sensors reported to date. Figure 2b demonstrates the detector's capability to detect photons down to energies of 550 eV by imaging the far field diffraction from a Fresnel Zone Plate focusing optic at this low energy. The Detective Quantum Efficiency (DQE) is estimated to be about 30% (see section 5) and logarithmic color scale emphasizes both the absence of noise and the large dynamic range.

## Performance of the system

Figure 3a shows the average calibrated pulse-height distributions measured for a shallow iLGAD sensor variation, with multiplication factor 4.8, with photon energies between 550 eV and 900 eV. Although detectable, photon energies below 550 eV are not shown since at these energies the Signal-to-Noise Ratio (SNR) and the DQE decreases considerably. Additionally, at lower energies the multiplication is mostly triggered by holes, resulting in a lower effective gain, since the majority of the photons convert before or in the gain layer. Consequently, signals from hole-triggered multiplication are often indistinguishable from noise. The main peak in the curves results from photons absorbed in the sensor bulk behind the gain layer undergoing electron-initiated multiplication with the totality of the charge collection in a single pixel. The flat region preceding the peak is caused by two factors: i) photons that convert at a shallower depth, which results in a lower charge multiplication factor, and ii) charge-sharing effects, where the generated charge is distributed across multiple readout pixels[52]. In the graph, the threshold level at with a Signal-to-Noise ratio of 5 is reached is highlighted, which is equal to 447 eV, for this variation.

Figure 3 b compares the average pulse-height distributions for standard and a shallow iLGAD sensors (with $M = 6.1$ and $M = 4.8$) respectively at 550 eV, 700 eV, and 900 eV. The standard iLGAD variation shows an increased flat region below the full-energy peak, as fewer photons convert within the thinner gain layer. The shift towards lower energies of the peak position, in the case of the standard variation, can be attributed to the same effect, as a smaller fraction of the events undergoes an optimal multiplication. Thanks to the higher $M$, a SNR of 5 is reached at a marginally lower threshold, 425 eV. However, a smaller fraction of photons will be detected, compared to the shallow variation. The ratio of photons undergoing electron-triggered multiplication to those absorbed before or within the gain layer influences the DQE of the detector. Charge-sharing effects are expected to be consistent across iLGAD variations, as they are influenced primarily by pixel size and charge collection time, which are determined by sensor thickness, bias voltage, and X-ray energy[53]. The calibration method and the procedure for obtaining the average pulse-height distributions shown in Fig. 3 are detailed in "Methods" section.

Figure 4a shows the noise as a function of the calibration gain $\mathcal{G}$ of the detector, which includes the combined gain of the readout electronics and the iLGAD multiplication factor $M$. The calibration gain is the conversion factor between signal amplitude and photon energy and it is expressed in mV/eV. $\mathcal{G}$ depends on the preamplifier gain settings of the ASIC and on the LGAD multiplication factor $M$ (see "Methods"). The noise is measured as Equivalent Noise Charge (ENC), expressed in electrons, which is the input signal required to produce an output equal to the root mean square (*rms*) noise. The gain and noise values are extracted for each pixel and averaged. The error bars indicate the standard deviation of the pixel distribution (see Supplementary Notes 1 and 2). A higher $\mathcal{G}$ is generally linked to a lower effective noise, as visible in Fig. 4a, and the use of LGADs significantly improves performance compared to an EIGER ASIC paired with a conventional silicon sensor under the same settings, as the signal is further amplified by the multiplication process in the sensor. A fit with a reciprocal function is added to the plot.

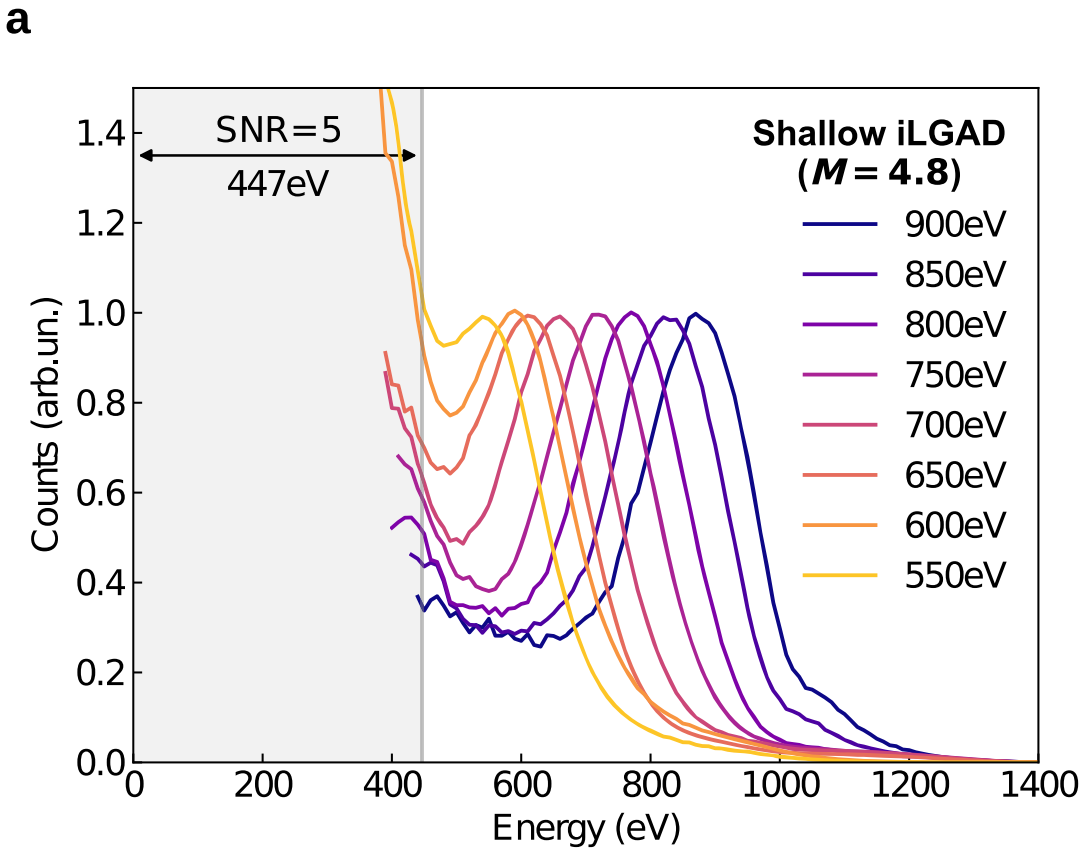

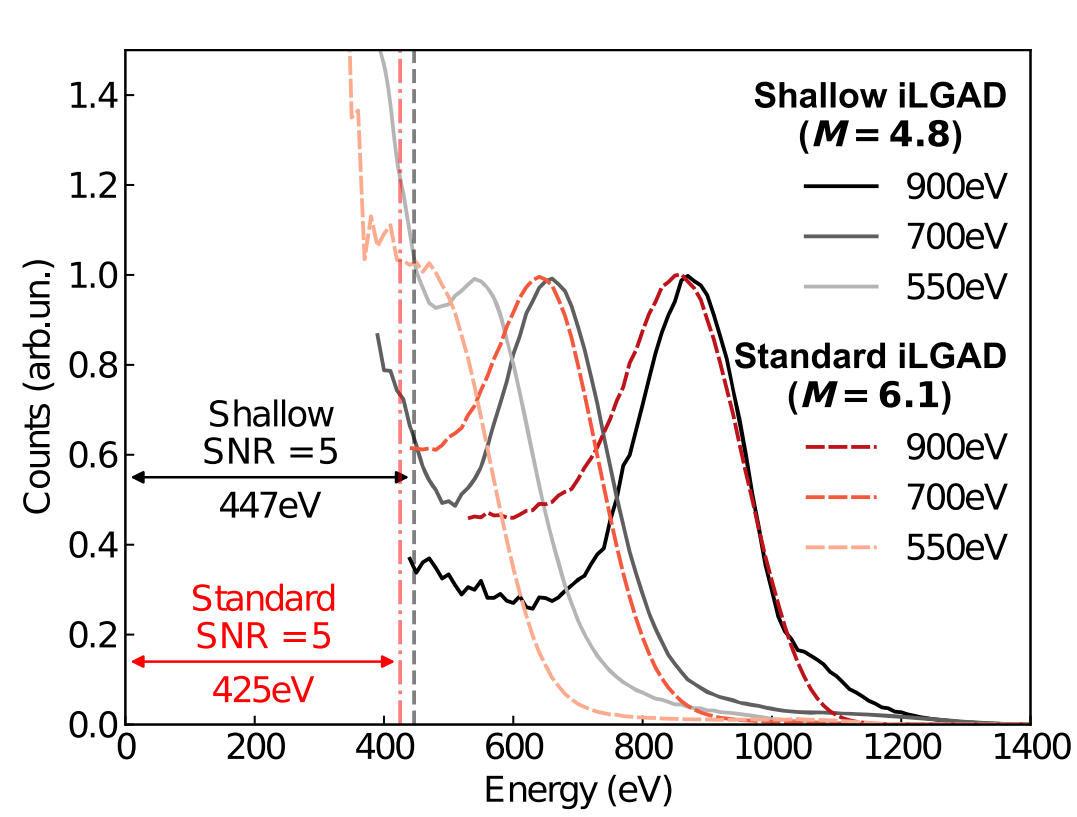

**Fig. 3 | Pulse-height distributions. a** Calibrated pulse-height distribution as a function of the X-ray photon energy for the shallow iLGAD variation, $M = 4.8$. The spectra are obtained as average after a pixel-wise calibration, and normalized to the peak integral. The gray band and horizontal arrow indicate the level at which a SNR of 5 is achieved, considering the ASIC settings optimized for 550 eV. The ASIC settings have been optimized for the individual energies. **b** Comparison of the calibrated pulse-height distributions obtained using a standard and a shallow iLGAD sensor variation at 900 eV, 700 eV, and 550 eV. The ASIC settings have been optimized for the individual energies and iLGAD sensor variations. Vertical lines indicate the levels at which a SNR = 5 is reached for each of the two sensor variations, considering the ASIC settings used for 550 eV. The curves have been normalized at the peak integral.

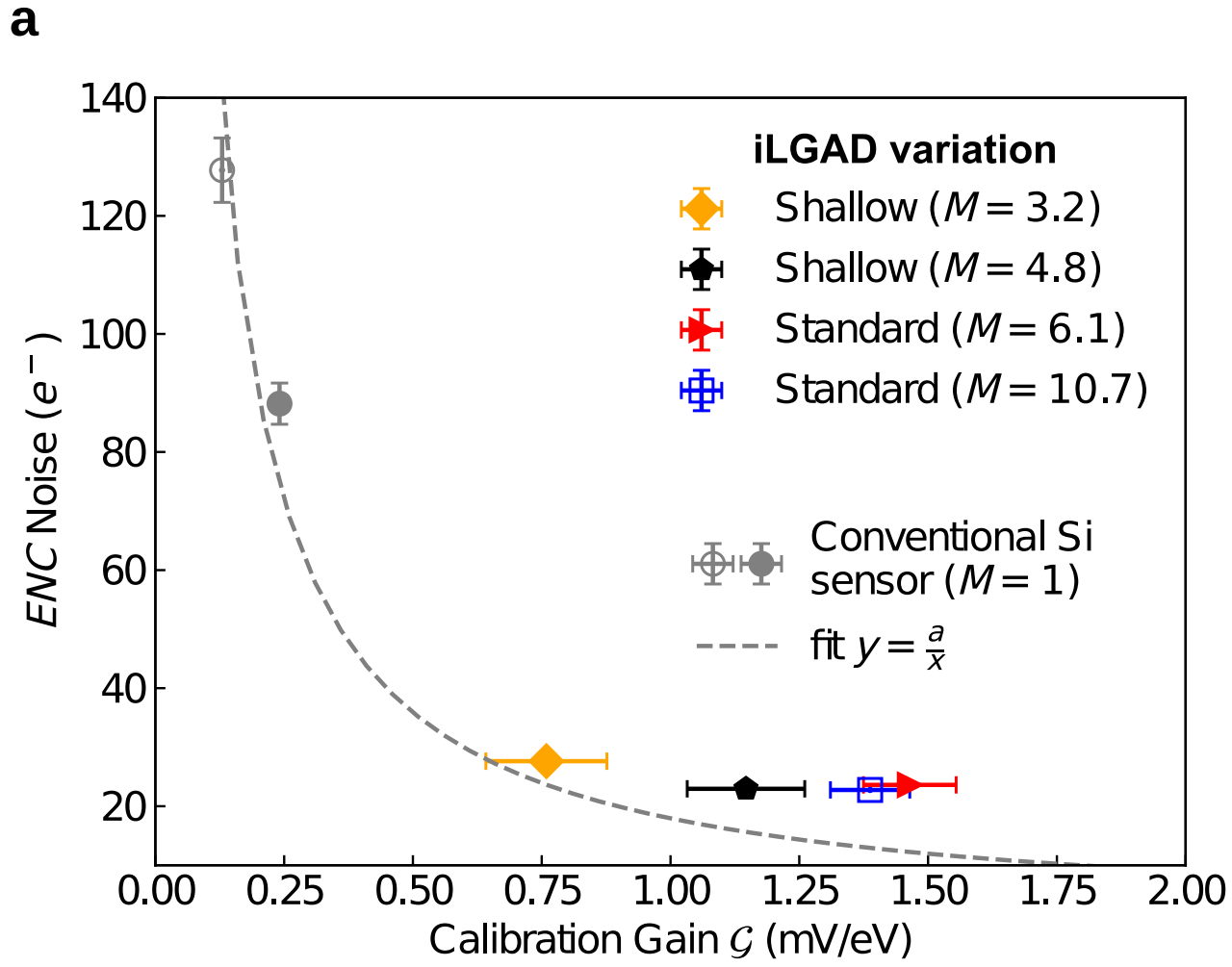

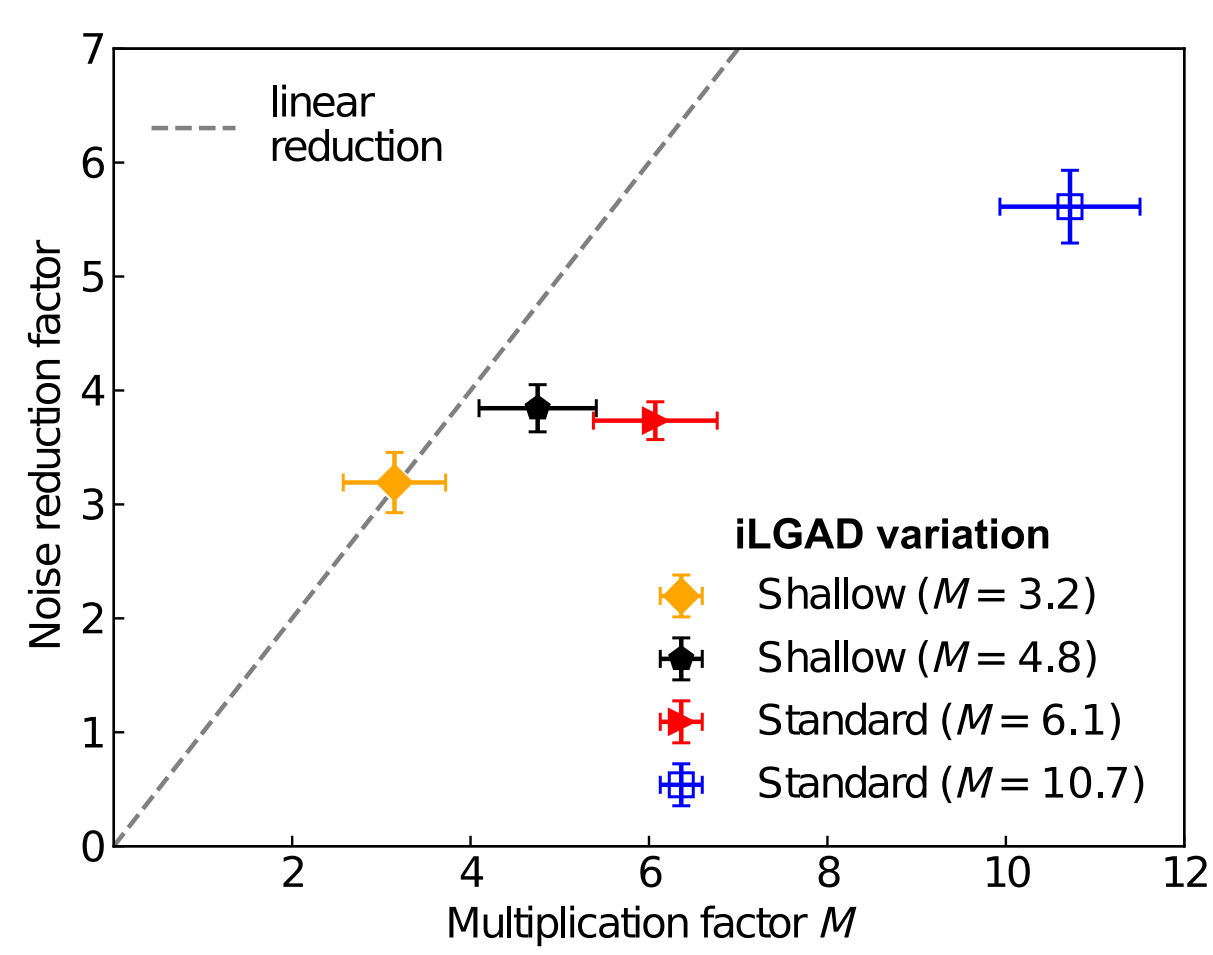

**Fig. 4 | Noise performance. a** Effective noise as a function of calibration gain $\mathcal{G}$, of four iLGADs samples of both standard and shallow variations, compared to a conventional silicon sensor. The data points are fitted with a function $y = a/x$, as the Effective Noise is expected to have an inverse relation with the calibration gain $\mathcal{G}$. For the data points two different settings of the EIGER ASIC are used, differentiated in the plot by filled or empty markers. **b** Noise reduction factor as a function of the LGAD multiplication factor $M$, with respect to a conventional silicon sensor, for the four iLGADS samples of both variations. The dashed line indicates the ideal case where the noise reduction scales linearly with $M$; the data points fall below this trend and are tabulated in the Supplementary Table 1. In both plots, the error bars indicate the standard deviation of the pixel distribution.

Ideally, the noise reduction of an LGAD sensor with respect to a conventional Si sensor scales linearly with $M$ when the same ASIC settings are used. For instance, with an LGAD sensor providing a $M \sim 3$, the effective noise is reduced from $90e^{-}rms$ to less than $30e^{-}rms$. However, with higher $M$, the noise reduction becomes less efficient, due to factors such as the excess noise from the multiplication process and the increased leakage current[54]. In this work, effective noise levels as low as $23e^{-}rms$ were achieved. As shown in Fig. 4b, for $M$ lower than 4, the noise is effectively reduced by a similar factor. However, as $M$ increases, the noise improvement becomes sub-linear. For instance, at a gain of around 5, the noise reduction is approximately a factor of 4, whereas for $M \sim 11$, the noise is reduced by less than a factor of 6. There is no noticeable difference in noise reduction between the shallow and standard iLGAD designs. However, a thicker gain layer typically enables a higher multiplication factor.

The SNR as a function of photon energy is shown in Fig. 5a. The average of the SNR calculated for each individual pixel is shown with the error bars indicating the standard deviation of the distribution. Both standard and shallow sensor variations achieve an SNR above 10 at about 850 eV. At 550 eV, the estimated SNR values are about 5.4 for both variations. The SNR varies only marginally between the two variations over the explored energy range. The optimal performance of a single photon counting detector is obtained for an SNR ≥ 10[55] and the threshold set at half of the photon energy. An optimal threshold at half of the photon energy, and SNR ≥ 5, can therefore be set only above 800 eV. While the detector can operate with a lower SNR, this compromises the Counting Efficiency (CE) because the threshold must be set higher than half the photon energy. Additional details are provided in the Supplementary Note 2.

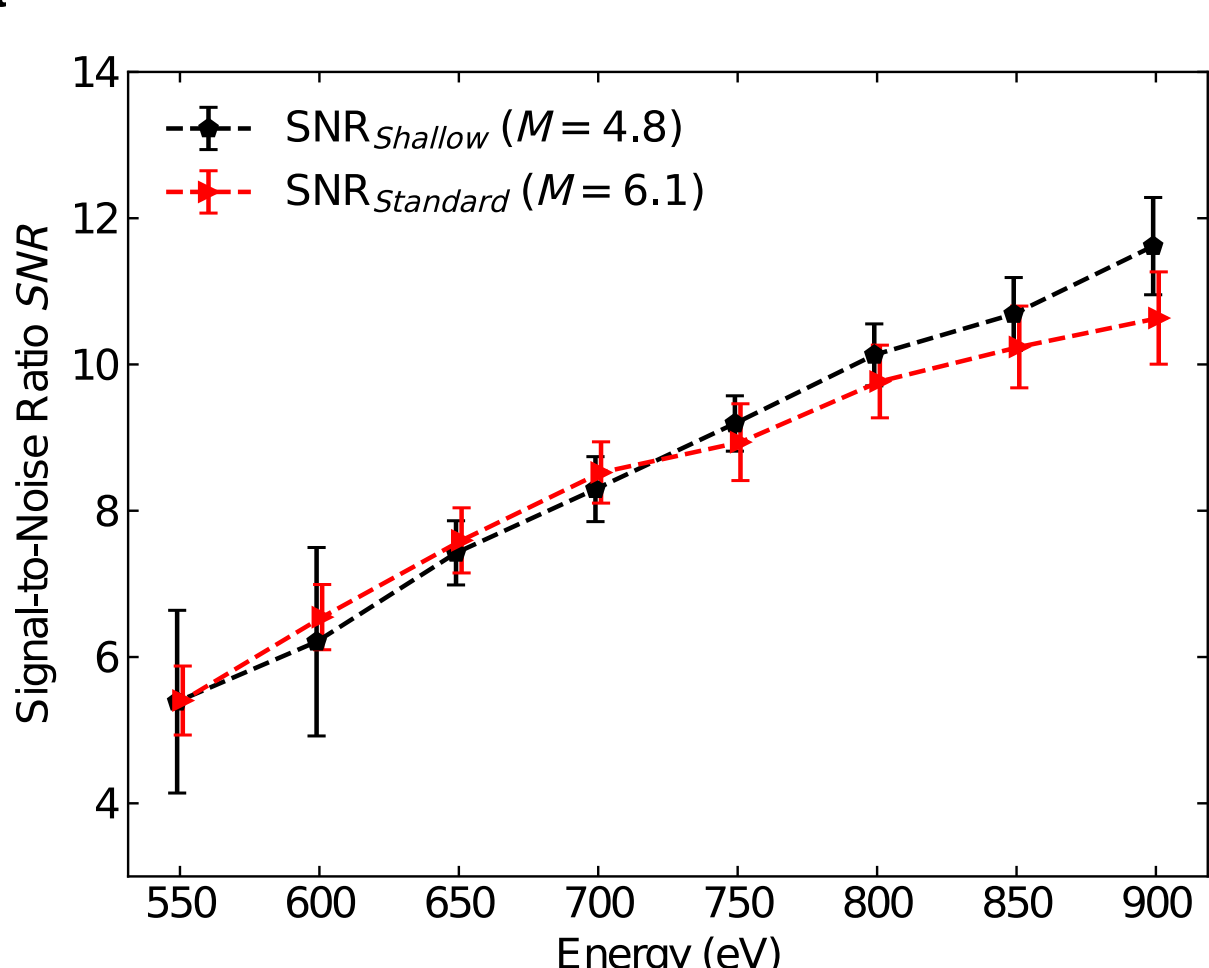

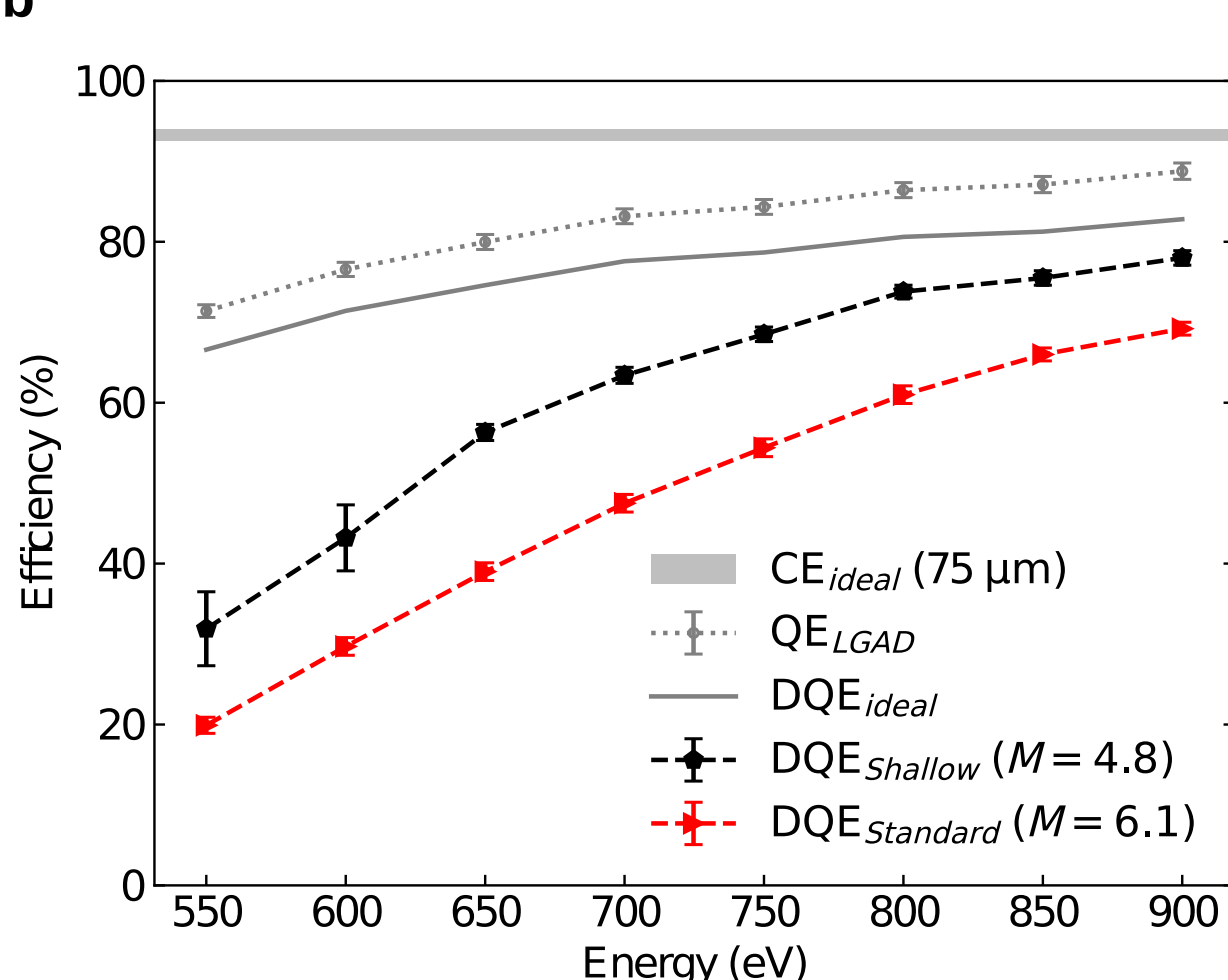

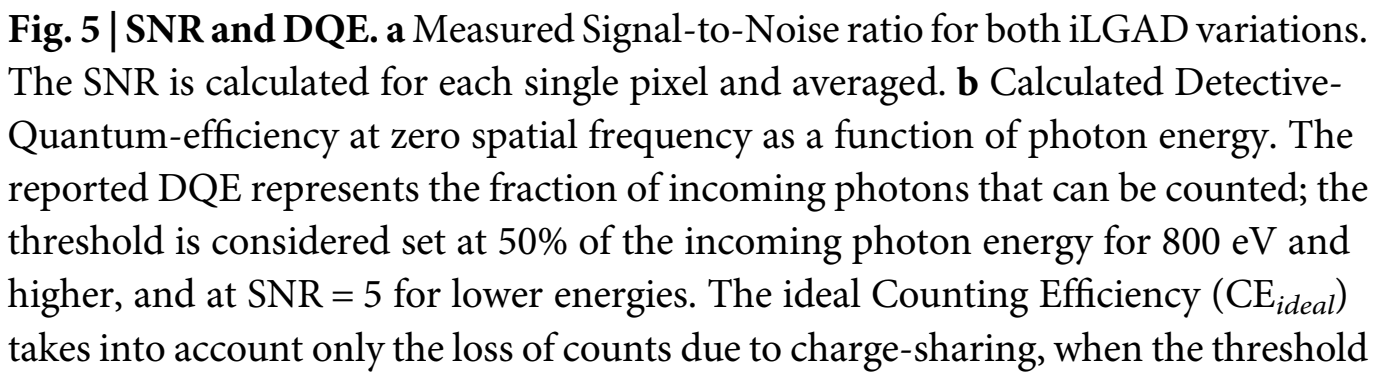

**Fig. 5 | SNR and DQE. a** Measured Signal-to-Noise ratio for both iLGAD variations. The SNR is calculated for each single pixel and averaged. **b** Calculated Detective-Quantum-efficiency at zero spatial frequency as a function of photon energy. The reported DQE represents the fraction of incoming photons that can be counted; the threshold is considered set at 50% of the incoming photon energy for 800 eV and higher, and at SNR = 5 for lower energies. The ideal Counting Efficiency ($CE_{ideal}$) takes into account only the loss of counts due to charge-sharing, when the threshold is set at half of the incoming energy, considering an ideal QE and noise-less electronics. $QE_{LGAD}$ is the measured Quantum Efficiency[45]. $DQE_{ideal}$ is the product between $CE_{ideal}$ and $QE_{LGAD}$, i.e., the ideal Detective Quantum efficiency achievable by an LGAD sensor, without losses due to the multiplication process and considering a noiseless electronics. In both plots, the error bars indicate the standard deviation of the pixel distribution. Additional details on the SNR estimation and DQE calculation can be found in the Supplementary Notes 1 and 2.

Figure 5b shows the calculated DQE of the detector as a function of photon energy, at zero spatial frequency. The DQE combines the QE of the sensor and the CE with the threshold set at the greater of either half the incoming energy or the energy at which SNR=5. It is a key figure of merit that describes how effectively a photon is converted into useful information in the final image. The calculation is based on the iLGAD parameters, measured in ref. 45, and on the measured noise levels and multiplication values from this work. The ideal Counting Efficiency $CE_{ideal}$ is defined as the fraction of total incident photons detectable with a threshold at half of the photon energy, assuming a noiseless electronics and 100% QE, it depends on the charge sharing and is only affected by photon loss in the pixel corners[56]. For a detector with pixel pitch as the ones used in this study, the maximum CE is ~93%; the charge-sharing effect has been estimated from the average pulse-height distributions in Fig. 2, with a procedure reported in Supplementary Note 3.

For SNR > 10, i.e., down to 800 eV, the DQE of the detector is close to the ideal value, given by the product between the QE of the sensor and the ideal CE, indicated as $DQE_{ideal}$ in Fig. 5b. The DQE decreases with the energy, due to the lower QE of the sensor, as well as to a reduced counting efficiency. This reduction in CE arises from the inability to set the threshold at half the photon energy because of noise constraints. Additionally, at low energies, a higher proportion of photons are absorbed before or within the gain layer. Consequently, the shallow variation with its thinner and shallower gain layer achieves a higher DQE over the whole energy range.

### Ptychographic imaging

Ptychography is a scanning CDI method that has been highly successful with hard X-rays, with resolution far surpassing the pixel size of the detector and the beam spot size limitations of direct imaging[16]. Single photon counting detectors are routinely used for recording diffraction patterns in far-field geometry, providing the high dynamic range needed to detect both the strong signals from the central undiffracted light cone and the weak signals from scattered photons at high Q-values. A fast frame rate is essential to match the high flux provided by modern X-ray sources and to facilitate rapid scanning[57]. Moreover, burst ptychography using a multi-kHz framing detector promises to further improve the resolution limits of this technique by reducing the effects of mechanical instabilities[51]. We focus on ptychography because, along with diffraction, it is one of the principal experimental techniques that has significantly benefited from single photon counting pixel detectors in the hard X-ray regime. In contrast, due to the stringent spatial resolution requirements, TXM primarily relies on CCD and CMOS sensors, while STXM does not benefit from the use of position sensitive detectors and usually relies on a point detector. The development of the EIGER-iLGAD detector now enables the extension of high-throughput, high-resolution ptychographic imaging into the soft X-ray energy range.

An EIGER-iLGAD detector with shallow sensor variation was installed at the Surface/Interface Microscopy (SIM-X11MA) beamline of the Swiss Light Source (SLS), Paul Scherrer Institut, Villigen, Switzerland[58] for commissioning and proof-of-principle experiments until the facility shut down for the upgrade to SLS 2.0 in September 2023. The detector was then installed in the SOPHIE endstation, which went into operation at the SoftiMAX beamline of the MAX IV synchrotron in Lund (Sweden), where it was available to the synchrotron user community until the end of June 2025.

In order to demonstrate the suitability of the detector for ptychographic imaging, a Siemens star imaged by soft X-ray ptychography at 707 eV, where the DQE of the detector is ~50%. The Siemens star was fabricated on a $Si_3N_4$ membrane by electron beam lithography using negative-tone hydrogen silsesquioxane (HSQ) resist. The developed HSQ resist was then coated with a 10 nm Ir film by atomic layer deposition (ALD). Details about the lithographical patterning method can be found in ref. 59. The diffraction patterns for the ptychographic scan were obtained with an exposure time of 2.5 ms. The comparatively large pixel size of 75 µm does not represent a hindrance for ptychography, as the sampling criterion of diffraction patterns in regular CDI is not as stringent for this technique[60,61].

The resulting amplitude and phase images are displayed in Fig. 6a and b, respectively. The spatial resolution achieved in this image was estimated using Fourier Ring Correlation (FRC) between two identical ptychographic scans[62], suggesting a resolution of 6.5 nm with a 1 bit threshold.

Figure 6c demonstrates the detector's performance in an X-ray linear dichroism (XLD) image of a freestanding $BiFeO_3$ (001) film. Additional images of the multiferroic domains structure in freestanding $BiFeO_3$, acquired with the detector presented here, have been published in refs. 63,64. The results in ref. 64 demonstrate that the EIGER-iLGAD detector is also capable of operation at the O K-edge (530 eV), even below 550 eV. However, while it remains functional at these lower energies, there is a reduction in DQE, combined with a lower X-ray scattering cross section from the ferroelectric domains of BFO, resulting in a lower spatial resolution in the reconstructed image. These results were obtained with a longer exposure time of 200 ms, which ensures high spatial resolution. Soft X-ray ptychography with the LGAD EIGER has also been employed to study magnetization dynamics in permalloy ($Ni_{81}Fe_{19}$) microstructures by XMCD imaging in pump-probe mode[65].

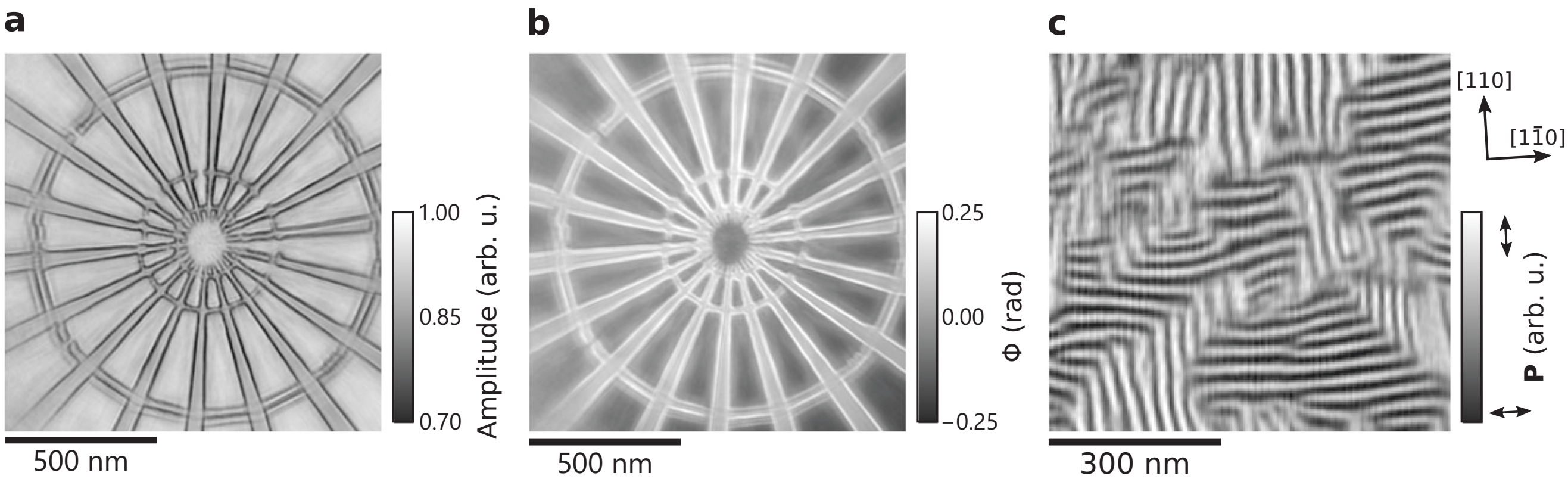

**Fig. 6 | Ptychographic images. a** Amplitude and **b** Phase ptychographic images of a Siemens star. These images were obtained with an exposure time of 2.5 ms. **c** X-ray linear dichroism image (phase contrast) showing the in-plane projection of the ferroelectric polarisation (**P**). Mosaic-like multiferroic domains are visible, which contain spin cycloids (64 nm period) in an 80 nm thin freestanding $BiFeO_3$ (001) film. The crystallographic axes of the $BiFeO_3$ film are indicated above the color bar. The contrast indicates both the in-plane axis of $P$ in the $BiFeO_3$ film and delineates the spin cycloid coupled within the multiferroic domains (see refs. 63,64 for details). All images were acquired with an X-ray energy of ~707 eV (Fe $L_3$-edge). At this energy, the maximum of the XLD contrast originating from the BFO ferroelectric domains is found in the phase information.

## Discussion

The presented system, combining the EIGER single-photon counting pixel ASIC with custom-developed LGAD sensors, successfully extends hybrid detector capabilities into the soft X-ray energy range, pushing the lower limit for single-photon resolution down to ~550 eV. This system achieves a SNR above 5, without compromising performance in frame rate, dynamic range, or noise-free operation, that are intrinsic of the photon-counting architecture. The energy range of the presented detector can be even pushed down to lower energies, by sacrificing its DQE. In general, the shallow iLGAD design offers higher DQE across all X-ray energies. Further increasing the $M$ while maintaining low noise performance could be beneficial for extending detection to even lower energies.

Despite the promising results, further developments are needed before this detector can fully meet the requirements of the entire soft X-ray imaging community. Improvements in both pixel and sensor yield are essential. Currently, the pixel yield is around 97%, with significant variability observed between different sensors, likely due to non-uniformity of the multiplication factor, as discussed in ref. 66. The pixel yield declines rapidly at higher temperatures and lower bias voltages, suggesting that leakage current may be causing saturation in the analog chain. Reducing the leakage current through sensor technology advancements would enable detection at lower photon energies and reduce effective noise. Enhanced cooling and optimized biasing could similarly improve performance. Additionally, the fabrication of large-area iLGAD of $4 \times 4$ $cm^2$ with high production yield still has to be demonstrated, considering that the first prototyping batch showed a yield limited to about 30%. A higher yield is necessary to reliably build multi-megapixel detectors, as is standard practice with conventional silicon sensors for hard X-rays.

Additional improvements can be achieved through advances in both readout electronics and iLGAD sensor technology. While reducing noise in single photon counting detectors to improve the minimum detectable energy is challenging, and likely limited to a 30-50% improvement in the short term, significant progress is being made to increase count-rate capabilities[67], a crucial step for operating single photon counting detectors at diffraction-limited light sources currently under construction worldwide. The high gain provided by LGADs enables faster signal shaping, which is essential for achieving the required count rate capabilities.

In the future, even lower energies could be achievable through the development of a lower-noise photon-counting ASIC paired with LGAD sensors. Simultaneously, advancements in LGAD technology are underway to increase the $M$ up to 20, enabling detection at lower energies and improving sensor QE by further thinning the entrance window. From the sensor perspective, the QE can be enhanced by reducing the passivation layer on the entrance window, while a higher counting efficiency may be achieved with even shallower and thinner gain implants. However, increasing $M$ to access lower energies is effective only if the leakage current, which raises shot noise and decreases pixel yield, remains controlled.

Single photon counting LGAD detectors will be instrumental for applications at soft and tender X-ray beamlines with the goal of extending to the EUV energy range with sensors with higher multiplication factors. Moreover, iLGAD sensors combined with charge-integrating readout find applications at X-ray Free Electron Lasers[66]. The possibility of applying position interpolation methods opens perspectives also for high resolution imaging, including Resonant Inelastic X-ray Scattering (RIXS), where single photon detection combined with high spatial resolution is essential and high frame rate detectors are lacking. The advancement of single photon counting detectors for soft X-rays enables new applications in low-energy photon science, offering high sensitivity and precision that promise to expand capabilities in fields from biological microscopy to materials science.

## Methods

### iLGAD sensors

Low Gain Avalanche Diodes (LGADs) are silicon-based sensors that include an additional moderately-doped implanted region (gain implant) of the same type of the substrate, where charge carriers are accelerated by a strong electric field (of the order of 300 kV/cm), initiating impact ionization and achieving a charge multiplication of approximately a factor of 10. The multiplication factor is mostly determined by the doping profile and concentration of the gain implant. The original LGADs developed for high-energy physics[43] feature a coarse segmentation, with about 1 mm pitch pixels and the gain implant matching the pixel layout. To prevent breakdown at pixel edges, termination structures are needed to create gain-free regions, which effectively reduces the fill-factor i.e., the proportion of the pixel where the charge produced by X-rays undergoes multiplication[68]. Various LGAD technologies aim to minimize or eliminate this limitation [69]. For the soft X-ray sensor presented here, an inverse-LGAD (iLGAD) design was used, featuring a uniform gain layer on the entrance window of the sensor[40]. The sensors used in this study are 275 μm thick. The sensors are fabricated using $p$ silicon wafers, with $p^+$ pixel implants. A $n^+$ implant on the entrance window side forms the collecting junction, and the additional $p^+$-type gain implant is formed just underneath the junction. Such a gain implant is unsegmented and it extends through the full sensor area, as shown in Fig. 1b. The sensors operate with hole collection, where holes drift toward the pixel while electrons move in the opposite direction (see Fig. 1b). The sensors were operated with a bias voltage of 300 V, although due to possible voltage drops along the bias line, the effective voltage reaching the sensors may be lower. The bias voltage has a small effect on the multiplication factor, while it can affect charge sharing.

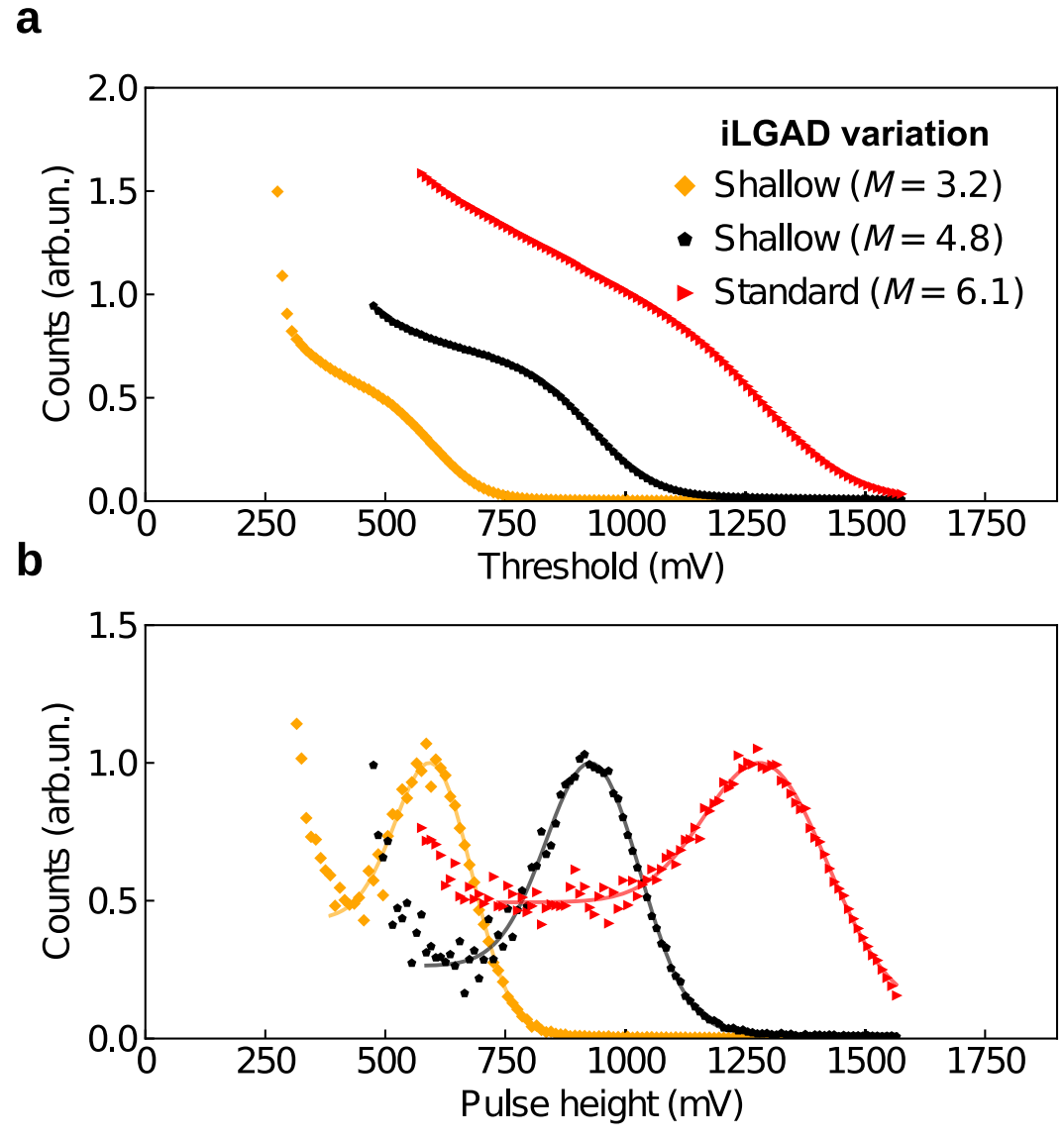

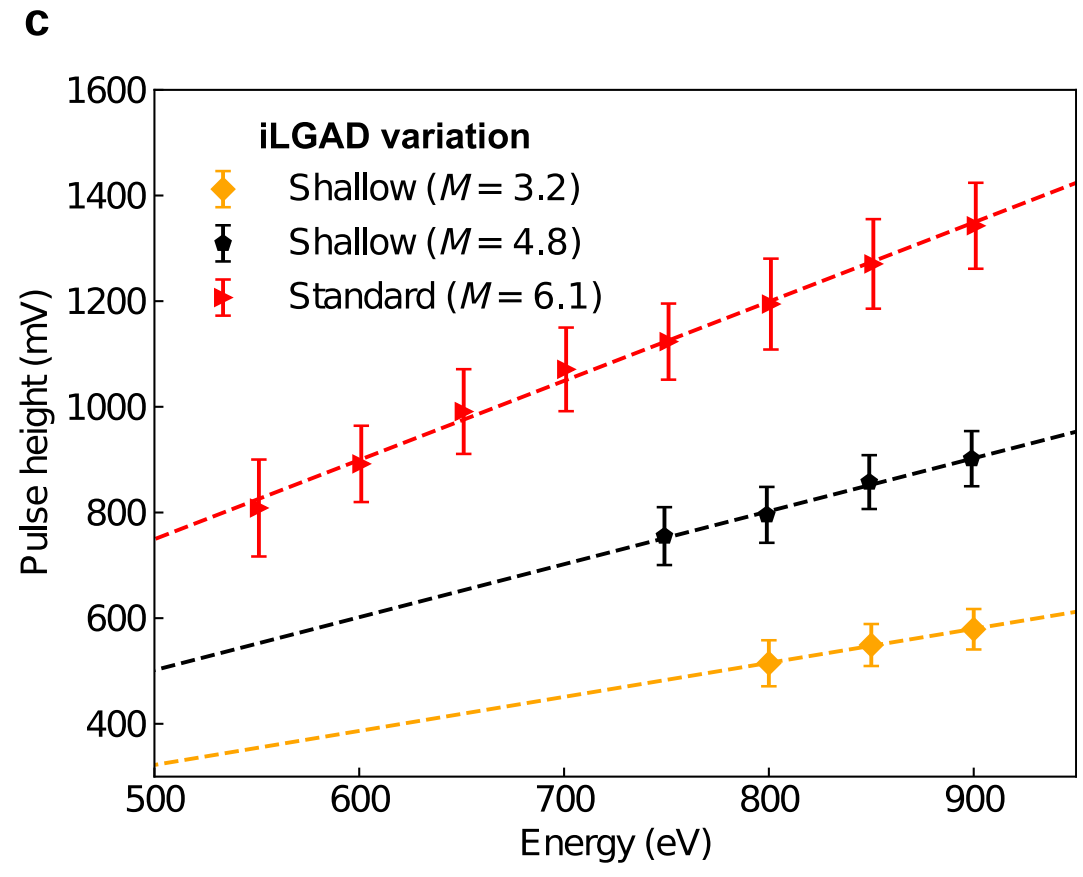

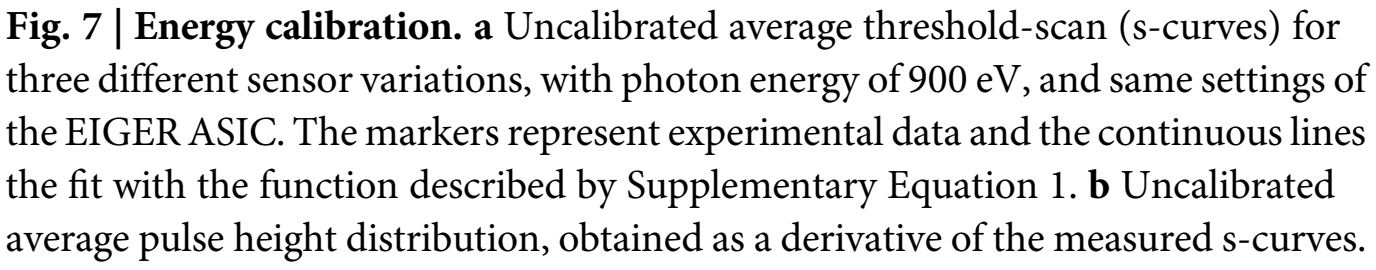

**Fig. 7 | Energy calibration. a** Uncalibrated average threshold-scan (s-curves) for three different sensor variations, with photon energy of 900 eV, and same settings of the EIGER ASIC. The markers represent experimental data and the continuous lines the fit with the function described by Supplementary Equation 1. **b** Uncalibrated average pulse height distribution, obtained as a derivative of the measured s-curves. **c** Correlation between photon energy and signal amplitude for three iLGAD variations of different $M$; the values are calculated as average over multiple pixels and the error bars represent their standard deviation; the energy calibration is performed via a linear fit (dashed lines), where the slope is $\mathcal{G}$.

The impact ionization can be triggered by electrons, when a photon is absorbed after the gain layer in the $p$-type bulk of the sensor or by the holes if a photon is absorbed before the gain layer. A mix of these two processes occurs when a photon is absorbed within the gain layer. Holes have a multiplication factor that is 2–4 times lower than electrons, due to their lower impact ionization coefficient, depending on the electric field and the doping profile of the gain layer. The multiplication factor values $M$ for iLGAD sensors refer specifically to electron-triggered multiplication[45]. The standard iLGAD variation has a gain layer thickness and doping profile similar to LGADs designed for tracking applications, while the shallow variation features a thinner gain layer closer to the surface to reduce hole-initiated multiplication. The iLGAD sensors presented here have an entrance window optimized for soft X-rays, featuring a thin $SiO_2$ and $Si_3N_4$ passivation layer for a total of about 80 nm instead of the aluminum layer used in conventional silicon sensors. Additionally, the doping species, profile, and concentration in the $n^+$-implant have been optimized. A new batch of sensors with even thinner passivation is under development, which is expected to further enhance the QE.

### EIGER single-photon counting read-out

The EIGER is a photon-counting ASIC, developed at PSI for diffraction experiments at synchrotron light sources. It features a 75 μm pixel pitch, with each ASIC comprising $256 \times 256$ pixels and covering an area of $2 \times 2$ cm$^2$[38]. Each pixel includes a low-noise charge preamplifier and shaper with adjustable gain and speed, a comparator with a globally set threshold, a 6-bit DAC (trim bits) to correct pixel-by-pixel inhomogeneities, and a 12-bit counter. The comparator in each pixel is triggered whenever the shaper signal exceeds the threshold, counting events in the pixel counter, which is read out after the exposure time. The gain of the analog chain is controlled by adjusting the preamplifier feedback resistance, set through a gate voltage. This adjustment allows a trade-off between noise (as low as ~$100e^-$ ENC *rms*[70]) and count-rate capability (up to ~1 Mcounts/pixel/s[71]). The low noise settings used in this work for low energy detection allow to achieve a count rate of the order of ~100kcounts/pixel/s. The 12-bit counter can be extended to 32-bits in firmware with minimal dead time between sub-frames, and the bit depth can be adjusted to reach frame rates up to 22 kHz in 4-bit mode.

In this study, the detector is built by arranging $2 \times 2$ ASICs ($4 \times 4$ cm$^2$, $512 \times 512$ pixels) bump-bonded to a single sensor of the same area. A 150 μm gap between adjacent ASICs creates a cross-shaped blind area, visible in Fig. 2a. Additionally, along the sensor edges, the outer 9 pixels are replaced by guard-ring structures to prevent breakdowns, making them blind. In this work, the threshold voltage is set uniformly across all four ASICs in the detector system, and the trim-bits correction is not used. The sensor and electronics are operated in vacuum at pressures lower than $10^{-5}$ mbar and at −28 °C, using a liquid chiller. Lower temperatures help reduce sensor leakage current and increase the LGAD multiplication factor, thereby enhancing the SNR[39,66].

The detector is connected to the readout board via flat-band cables and a vacuum flange patch panel, limiting the readout clock speed to half of its nominal value and capping the maximum frame rate at 10 kHz.

### Calibration

The characterization and calibration of the detector are performed via threshold scans, where counts are measured as a function of the comparator threshold using a constant-flux monochromatic photon beam. An example of the resulting s-curves is shown in Fig. 7a. The uncalibrated pulse-height distributions shown in Fig. 7b are obtained as the derivative of the threshold scans. Various iLGAD variations with different multiplication factor $M$, at 900 eV, averaged over a large number of pixels are compared. The preamplifier settings of the EIGER ASIC are the same for all the datasets.

The threshold scans can be fitted on a pixel-by-pixel basis using an s-curve function, as described in ref. 72. This fit enables extraction of the photon count, full-energy peak position, noise level, and fraction of shared charge. Pixel-wise calibration is performed with a linear fit of the peak position in the pulse-height distribution as a function of the beam energy. Figure 7c shows the position of the full-energy peak as a function of the photon energy, with linear fits; the data points represent the average and the error bars the standard deviation of the distribution of the pixels. The calibration gain of the detector $\mathcal{G}$ is defined by the slope obtained from this fit and depends on both the gain settings of the pixel preamplifier and the multiplication factor of the LGAD sensor $M$.

The noise value, derived from the s-curve fit is converted using the calibration gain $\mathcal{G}$. It represents the standard deviation of the calibrated photon peak and is expressed in units of electron-hole pairs, which correspond to 3.6 eV per pair in silicon. With a fixed $M$, a higher $\mathcal{G}$ generally yields a lower noise level (see Fig. 4a), though it comes at the expense of count-rate capability. The $M$ of an LGAD sensor can be estimated as the ratio of the average conversion gains to that measured with a conventional silicon sensor, under the same preamplifier gain settings (see Supplementary Note 1).

### Ptychography

Ptychographic imaging with the EIGER-iLGAD detector was performed using the SOPHIE (Soft X-ray Ptychography Highly Integrated Endstation) endstation. The sample to detector distance was 96 mm. The X-ray beam, tuned to the Fe $L_3$-edge (707 eV), was focused to a 400 nm FWHM spot on the sample using a 500 μm diameter line doubled Ir Fresnel zone plate with 50 nm outer zone width. A step size of 50 nm was chosen for the ptychographic scan on a Fermat spiral trajectory. The ptychographic scan for the Siemens star in Fig. 6a, b was carried out with an exposure time of 2.5 ms, which led to a total scan duration of 15 s with a dwell time of 22.5 ms. The XLD phase image of the $BiFeO_3$ film shown in Fig. 6c was acquired with a 220 ms dwell time for each position, resulting in a total acquisition time of ~3 min for a $1.5 \times 1.5\ \mu m^2$ scan area. A 500 μm diameter line doubled Ir zone plate with an outer zone width of 20 nm was used for imaging the multiferroic domains in the BFO thin film.

The reconstruction of the images was performed with 1200 iterations of difference-map[73] and 200 iterations of maximum-likelihood refinement[74] implemented in the Ptychoshelves software package[75]. The illuminating wavefront was reconstructed with three probe modes[76].The XLD phase image in Fig. 6c was computed by subtracting the drift-corrected phase images obtained with horizontal and vertical linear X-ray polarization.

## Data availability

All the data that support the findings of this study are publicly available in the following repository: [http://doi.psi.ch/detail/10.16907%2Fc0b2168f-d6de-4e3d-bd96-b20a9b505944]. Scripts to access the data, as well as examples of analysis, are available in the repository: [https://doi.org/10.5281/zenodo.15740655].

## Acknowledgements

Measurements with soft X-rays were performed at the Surface/Interface Microscopy (SIM-X11MA) beamline of the Swiss Light Source (SLS), Paul Scherrer Institut, Villigen, Switzerland. Additionally, soft X-ray ptychography measurements were carried out with the SOPHIE endstation installed at the SoftiMAX beamline of the MAX IV Laboratory. The SOPHIE endstation was designed and assembled at the SLS, PSI, Villigen, Switzerland. Research conducted at MAX IV, a Swedish national user facility, is supported by the Swedish Research council under contract 2018-07152, the Swedish Governmental Agency for Innovation Systems under contract 2018-04969, and Formas under contract 2019-02496. T.A.B. acknowledges funding from the Swiss Nanoscience Institute (SNI) and the European Regional Development Fund (ERDF). N.W.P. received funding from the European Union's Horizon 2020 research and innovation programme under the Marie Skłodowska-Curie grant agreement no. 884104. We thank Benedikt Rösner for the fabrication of the Siemens star, Chia-Chun Wei, Shih-Wen Huang and Jan-Chi Yang for the fabrication and characterization of the freestanding $BiFeO_3$ film. We thanks Armin Kleibert and Carlos Vaz for the support at the SIM beamline of the Swiss Light Source. We additionally thank Karina Thånell and Igor Beinik for support at the SoftiMAX beamline at Max IV.

## Author contributions

Conceptualization: F.B. and A.B. Methodology: F.B., A.B., M. Bo., M. Br., M.C., M.C.V., R.D., D.G., E.F., A.M., G.P., and J.Z. Investigation: F.B., A.B., T.A.B., S.F., E.F., and N.W.P. Visualization: F.B., T.A.B., and S.F. Funding acquisition: A.B. and B.S. Project administration: A.B., B.S. Supervision: A.B., J.R. and B.S.

## Competing interests

The authors declare no competing interests.

## Additional information

**Supplementary information** The online version contains supplementary material available at https://doi.org/10.1038/s42005-025-02240-9.

**Correspondence** and requests for materials should be addressed to Anna Bergamaschi.

**Peer review information** *Communications Physics* thanks the anonymous reviewers for their contribution to the peer review of this work.